# *Spin freezing and Field induced transition in $(Tb_{1-x}Eu_x)_2Ti_2O_7$: A Magnetic Property study*

Prajyoti Singh[1], Arkadeb Pal[1], Vinod K. Gangwar[1], Surajit Ghosh[1], Ranjan K. Singh[2],

A. K. Ghosh[2] and Sandip Chatterjee[1#]

[1] Department of Physics, Indian Institute of Technology (Banaras Hindu University), Varanasi-221005, India

[2] Department of Physics, Banaras Hindu University, Varanasi-221005, India

#Corresponding author's email address: schatterji.app@iitbhu.ac.in

## *Abstract*

The structural, magnetic and Raman effect have been investigated on $(Tb_{1-x}Eu_x)_2Ti_2O_7$. From structural study it is clear that Eu substitutes the Tb in $Tb_2Ti_2O_7$. Raman effect study indicates the existence of hardening due to phonon-phonon anharmonic interaction. From dc-magnetic measurement it is observed that in Eu rich samples contribution of dipolar interaction is significant. From ac magnetic measurement of $(Tb_{1-x}Eu_x)_2Ti_2O_7$ a new single ion weak spin freezing ~ 33K at zero magnetic field is observed. On applying a field of 1T, all the compounds show a field induced transition ($T^*$) which shifts towards higher temperature with increase of Eu content. This field induced transition corresponds to single moment saturation.

## *Introduction*

Geometrically Frustrated pyrochlore systems $A_2B_2O_6O'$ where A is a rare earth ion and B is transition metal ion [1], have attracted great attention both experimentally and theoretically in past decades owing to their different low temperature states [2-5]. In pyrochlore frustrated system, a balance between exchange, dipolar interactions, single ion anisotropy and strong crystal field effect lead to a formation of exotic degenerate ground states [6-8]. These ground states include spin liquid [6, 9, 10], spin glass [11-13], spin ice [7, 8, 14-16] and order by disorder [17-19]. In cubic pyrochlore oxide $R_2Ti_2O_7$ (space group Fd-3m), the $Ti^{+4}$ ions are non magnetic and the rare earth ion $R^{+3}$ which resides on a lattice of corner sharing tetrahedron has a key role to determine the magnetic properties of frustrated systems. In this geometry, each $R^{+3}$ ion is surrounded by 8 oxygen atoms showing a distorted trigonal cube of $D_{3d}$ symmetry which leads to frustration [20]. Strong crystal field interaction with the $D_{3d}$ symmetry result the differences between $R^{+3}$ ionic magnetic susceptibilities along $D_{3d}$ axis and its perpendicular direction which develop single ion anisotropy [21]. The pyrochlore structure with specific antiferromagnetic nearest neighbour exchange interaction shows no transition to long range magnetic ordering at finite temperature as its classical Heisenberg magnetic moment interact with a nearest neighbour [22, 23, 24, 25]. In such type of magnetic systems, it has been suggested that S=1/2 pyrochlore Heisenberg antiferromagnet is fully quantum disordered that they possess a spin liquid state [26] which have strong and non–trivial short range spin correlations. But some insulating magnetic systems show either long range ordering as in $FeF_3$ [27], $Gd_2Ti_2O_7$ [28, 29, 30], $ZnFe_2O_4$ [31] and $ZnCr_2O_4$ [32] or show spin glass phase in $Y_2Mo_2O_7$ [33,34,35], $Tb_2Mo_2O_7$ [33-36], $Y_2Mn_2O_7$ [37], $CsNiCrF_6$ [38] pyrochlore etc.

One of our parent compounds spin liquid $Tb_2Ti_2O_7$ (TTO) has ideal, disorder free structure as it is free from A/B disorder or oxygen nonstiochiometry [39] but some deviations in Tb/Ti ratio influence the spin lattice coupling below 1K [40]. Pyrochlore $Tb_2Ti_2O_7$ shows a negative Curie-Weiss temperature ($\theta_{CW}$=-14.75K) confirming its effective interaction between spins to be antiferromagnetic. Therefore, <111> anisotropy should not develop magnetic interaction and a long range magnetic ground state is expected, but in this case no such sign of long range magnetic ordering is found down to 50mK [41-44]. From different measurements like elastic and inelastic neutron scattering (INS), muon spin relaxation (µSR) it has been shown that this pyrochlore system develops AFM short range order at ~50K, but it remains in paramagnetic state down to temperature ~0.07K [41]. From recent study, it has been proposed that $Tb_2Ti_2O_7$ crystal possesses two phases at low temperature and low magnetic field applied along (111) direction [45]. In its magnetic field dependent ac susceptibility measurement, two weak magnetization plateaus were found at temperatures 16 mK and 40mK respectively which suggested that 1st phase to be Quantum spin ice and 2nd one to be quantum Kagome ice [45].

Another parent compound $Eu_2Ti_2O_7$ (ETO) is also an antiferromagnetic in nature ($\theta_{CW}$= -1.35K) [46]. From earlier study it has been observed that $Eu_2Ti_2O_7$ has temperature dependent anomalous correlation effect in $\chi^{-1}$ and specific heat measurement at low temperature [47]. It has been shown that this effect is due to the super-exchange interactions between $Eu^{+3}$ ions at intrinsic and defect states [47]. An interesting result was found in its high temperature magnetic susceptibility measurement as its ac susceptibility introduces a sharp spin freezing below 35K (~32K for 500Hz) which is thermally activated process. From Cole-Cole plot and Casimir–du Pre relation, it has been confirmed that this high temperature spin freezing is single ion relaxation process [46].

In this paper we report that $Tb_2Ti_2O_7$ shows a glass like transition below 36K (~33K for 500Hz) at zero magnetic field, observed from its imaginary part of ac susceptibility which is not reported till now. We have also investigated the structural, magnetic and Raman study of $(Tb_{1-x}Eu_x)_2Ti_2O_7$ (TETO) series over the full composition range between $Tb_2Ti_2O_7$ and $Eu_2Ti_2O_7$. Its DC magnetic study indicates the decrease in total magnetic moment on increasing the Eu concentration gradually. AC magnetic study reveals appearance of a new field induced peak at lower temperature on applying a field of 1T.

## *Experiment*

Polycrystalline powder samples of $(Tb_{1-x}Eu_x)_2Ti_2O_7$ with (x= 0,0.25,0.50,0.90,0.95,1.0) were synthesized using the solid state reaction method. High purity (99.999%) powder of $Tb_4O_7$, $Eu_2O_3$ and $TiO_2$ were taken in an appropriate stoichiometric ratio and ground for half an hour and then heated at $1000^0C$ in air for 24 hours. This calcination process was carried out several times. The resulting powder were pressed into pellets and sintered at $1400^0C$ for 48 hours. The prepared samples were characterized by X- ray powder diffraction (XRD) method at room temperature by using Rigaku Miniflex II X-ray diffractometer at wavelength of 1.5418 Å of CuKα radiation. All the dc and ac magnetic measurements were carried out using a Quantum Design magnetic property measurement system (MPMS) super conducting quantum interference device (SQUID) magnetometer down to 1.8 K. Raman spectra were recoreded by Renishaw Micro Raman Spectrometer using a solid state Laser of wavelength 532 nm.

## *Results and Discussions*

## *Stability Study*

Formation and stability of $R_2Ti_2O_7$ pyrochlore mainly depend on the ionic radius of $R^{3+}$ and $Ti^{4+}$ cations. If the ionic radius ratio $r(R^{3+})/r(Ti^{4+})$ lies between 1.46-1.78, then it forms a pyrochlore structure and if it is less than 1.46 then the system becomes a defect fluorite and if it is more than 1.78 then compound changes into pervoskite layered structure [48]. We have calculated the ionic radius ratio of all the samples of TETO series by using this equation [49]:

$$\frac{r(R^{3+})}{r(Ti^{4+})} = \frac{(1-x)r(Tb^{3+}) + xr(Eu^{3+})}{r(Ti^{4+})} \qquad (1)$$

Here we take the Shannon ionic radius of $Tb^{3+}$, $Eu^{3+}$, $Ti^{4+}$ (according to their Coordination number) ions which are 1.04Å, 1.066Å, 0.605Å respectively.

The calculated values are arranged in table 1. From this table it is clear that the all samples are shaped into a pyrochlore structure as the value of their ionic radius ratio lies in the range of 1.72-1.76. However the $r(R^{3+})/r(Ti^{4+})$ ratio increases with increase of $Eu^{3+}$ content which indicates the increased disorder in its structure.

| S.N. | Sample | $r(R^{3+})/r(Ti^{4+})$ |
|---|---|---|
| 1 | $Tb_2Ti_2O_7$ | 1.72 |
| 2 | $Tb_{1.5}Eu_{0.5}Ti_2O_7$ | 1.73 |
| 3 | $Tb_{1.0}Eu_{1.0}Ti_2O_7$ | 1.74 |
| 4 | $Tb_{0.2}Eu_{1.8}Ti_2O_7$ | 1.75 |
| 5 | $Tb_{0.1}Eu_{1.9}Ti_2O_7$ | 1.76 |
| 6 | $Eu_2Ti_2O_7$ | 1.762 |

Table (1): The ionic radius ratio of TETO series compounds.

## *Structural Study*

All the samples were examined by XRD measurement and it is found that all the samples were formed in single phase with no impurity. Fig. (1) shows the XRD pattern of all the samples. Inset of Fig. (1) represent the diffraction peak (222) of TETO series. From this feature it is clear that $Eu^{3+}$ ions replace the $Tb^{3+}$ ions moderately as diffraction peak shifts towards lower angle side ~~site~~ because the ionic radius of $Eu^{3+}$ ion is larger than that of the $Tb^{3+}$ ion. We have also extracted the structural parameters of all the samples from Rietveld refinement which was performed by FULLPROF program [50] and all its related information are listed in table 2. The Rietveld refinement diagrams of all the compositions are shown in Fig. (2). The lattice constants of all the samples quasi-linearly increase with $Eu^{+3}$ content as has been shown in Table (2).

| Parameters | X = 0 | X = 0.25 | X = 0.50 | X = 0.90 | X = 0.95 | X = 1.0 |
|---|---|---|---|---|---|---|
| Lattice Constant (Å) | a =b=c =10.154713 | a =b=c =10.162961 | a =b=c =10.175296 | a =b=c =10.191189 | a =b=c =10.19969 | a =b=c =10.20912 |
| Cell Volume ((Å$^3$) | 1047.13562 | 1049.6893 | 1053.4865 | 1058.4641 | 1061.0833 | 1064.05701 |
| Tb site X Y Z | 16d 0.5 0.5 0.5 | 16d 0.5 0.5 0.5 | 16d 0.5 0.5 0.5 | 16d 0.5 0.5 0.5 | 16d 0.5 0.5 0.5 | - - - - |
| Eu site X Y Z | – – – – | 16d 0.5 0.5 0.5 | 16d 0.5 0.5 0.5 | 16d 0.5 0.5 0.5 | 16d 0.5 0.5 0.5 | 16d 0.5 0.5 0.5 |
| Ti site X Y Z | 16c 0.0 0.0 0.0 | 16c 0.0 0.0 0.0 | 16c 0.0 0.0 0.0 | 16c 0.0 0.0 0.0 | 16c 0.0 0.0 0.0 | 16c 0.0 0.0 0.0 |
| O(1) site X Y Z | 48f 0.32928 0.125 0.125 | 48f 0.32192 0.125 0.125 | 48f 0.32428 0.125 0.125 | 48f 0.30553 0.125 0.125 | 48f 0.31993 0.125 0.125 | 48f 0.3363 0.125 0.125 |
| O(2) site X Y Z | 8b 0.375 0.375 0.375 | 8b 0.375 0.375 0.375 | 8b 0.375 0.375 0.375 | 8b 0.375 0.375 0.375 | 8b 0.375 0.375 0.375 | 8b 0.375 0.375 0.375 |
| $R_{wp}$ $R_{exp}$ $R_{wp}/R_{exp}$ $Chi^2$ | 15.7 12.4 1.26 **1.6** | 16.5 12.05 1.37 **1.87** | 17.3 13.4 1.29 **1.66** | 25.7 15.27 1.68 **2.84** | 22.4 18.12 1.23 **1.53** | 23.9 17.45 1.37 **1.88** |
| $d_{Eu-Eu}(Å)$ | -- | 3.59315 | 3.59748 | 3.60313 | 3.6061 | 3.60947 |

Table (2) : Different parameters obtained from Rietveld refinement of all TETO series compounds with Space Group F d -3 m .

## *Raman Study*

Fig.3 (a) shows the Raman spectra of TETO series at 300 K. From previous literature data [51-55] we assigned the bands. Phonon band near 512 cm$^{-1}$ is assigned to $A_{1g}$ mode. From the temperature dependent Raman study of $Tb_2Ti_2O_7$, it has been found that there are two peaks at 291 cm$^{-1}$ and 331 cm$^{-1}$ corresponding to $F_{2g}$ and $E_g$ band respectively at 4.2K but for room temperature this $E_g$ band becomes weaker [51-55]. Therefore, the most intense peak in the present investigation ~ 297cm$^{-1}$ can be assigned to a combination of two phonon modes [$F_{2g}+E_g$]. The fourth mode which has a weak signal at 451cm$^{-1}$ is assigned to $F_{2g}$ mode. Among other two $F_{2g}$ phonon modes, one which is observed ~209cm$^{-1}$ is visible in the present investigation and the other ~ 557cm$^{-1}$ is not visible. The another peak which is located near at 682cm$^{-1}$ represented as(*) is assigned to the combination mode as it is not a forbidden IR mode [56].

In fig. 3(b), we have shown the variation of phonon modes with the increasing concentration of Eu ions for all the samples. A blue shift i.e. hardening of modes is observed for the $F_{2g}+E_g$ ~ 297cm$^{-1}$ and for $F_{2g}$ ~ 209 cm$^{-1}$ band whereas for $F_{2g}$ ~ 451cm$^{-1}$ and $A_{1g}$ ~ 512 cm$^{-1}$ red shift i.e. softenening of mode is found. Generally The locations of Raman active bands depend on the bond strength, bond length and ionic masses .In our series the lattice parameter of compounds are increasing continuously on increasing the concentration of Eu ions and the ionic mass of the Eu ion is also less than Tb ion. Due to these factors the phonon should shift to lower frequencies for TETO series and this softening behaviour of phonons are shown by $A_{1g}$~ 512 cm$^{-1}$ and $F_{2g}$ ~ 451cm$^{-1}$ modes. But $F_{2g}+E_g$ ~ 297 cm$^{-1}$ mode and another $F_{2g}$ ~ 209 cm$^{-1}$ mode show a hardening behaviour. This hardening behaviour of phonon modes arises due to strong phonon – phonon anharmonic interaction in the system [57-58]. This gradual shifting in all the modes with increasing the Eu doping in Tb site confirms the substitution of Eu ion.

## *DC Magnetic Study*

In order to investigate the magnetic property of the TETO, we have measured the temperature and magnetic field dependent DC magnetization. Fig.4 (a) shows temperature dependent magnetization of all the samples at zero field cooled (ZFC) condition. This M-T curve reveals the paramagnetic behaviour for all the samples. The value of magnetic moment changes with Eu content as is clear from their M-T curves. Spin Liquid compound TTO possess highest magnetic moment among all the samples due to presence of magnetic $Tb^{3+}$ ion. When $Tb^{3+}$ ion is substituted by the $Eu^{3+}$ ion, the magnetic moment decreases and it becomes minimum for $Eu_2Ti_2O_7$. The reason behind the loss in magnetization is the continuous increment of $Eu^{3+}$ ion in place of $Tb^{3+}$ ion has higher magnetic moment compared to $Eu^{3+}$ ion. Interesting feature is observed in its inverse dc susceptibility ($\chi^{-1}$) Vs temperature graph which is shown in the inset of Fig.4(a). $\chi^{-1}$ for the first three samples in which $Tb^{3+}$ ions have equal or larger occupancy than the $Eu^{3+}$ ions i.e. $Tb_2Ti_2O_7$, $Tb_{1.5}Eu_{0.5}Ti_2O_7$, $Tb_{1.0}Eu_{1.0}Ti_2O_7$ show linear behaviour with temperature due to presence of strong crystal field interactions. On the other hand for the samples in which $Eu^{3+}$ ions have larger occupancy i.e. $Tb_{0.2}Eu_{1.8}Ti_2O_7$, $Tb_{0.1}Eu_{1.9}Ti_2O_7$, $Eu_2Ti_2O_7$ crystal field interactions are weak at low temperature (below 5 K) as their $\chi^{-1}$ sharply drop with decrease in temperature. This indicates the presence of other magnetic interactions *viz.* exchange, dipolar interactions etc. [59]. Due to this interesting observation we have analyzed the $\chi^{-1}$ data in three parts.

*(A)* We have fitted the $\chi^{-1}$ data of $Tb_2Ti_2O_7$, $Tb_{1.5}Eu_{0.5}Ti_2O_7$, $Tb_{1.0}Eu_{1.0}Ti_2O_7$ samples with normal Curie-Weiss (CW) Law

$$\chi^{-1} = \frac{T}{C} - \frac{\theta_{CW}}{C} \qquad (2)$$

Where $\theta_{CW}$ is Curie-Weiss Temperature and C is Curie Constant. From Fig.5(a), it can be seen that data fitted in (25-300K) temperature range. The theoretical effective magnetic moment of all the samples is calculated using the relation;

$$\mu^2_{eff} = x\mu^2_{Tb} + y\mu^2_{Eu} \qquad (3)$$

where x and y are number of Tb and Eu atoms per f.u. respectively. $\mu_{Tb}$ and $\mu_{Eu}$ are the effective magnetic moment of Tb and Eu ions respectively. From CW fit, the effective magnetic moment is calculated using $C = \frac{N\mu^2_{eff}}{3k}$ where N is Avogadro's No. and k is Boltzmann Constant. The resultant values obtained from the CW fit are listed in table (3). For $Tb_2Ti_2O_7$ the C.W. temperature is -14.75K and effective magnetic moment is $9.29\mu_B/Tb^{3+}$ ion which is consistent with earlier report [44]. The remaining two samples have more negative C.W. temperature indicating more antiferromagnetic interactions in those samples. The reason for that could be the increase in the effective distance between the Eu ions ($d_{Eu-Eu}$) which is listed in table (2).

| Sample | $\theta_{CW}$ | Theoretical $\mu_{eff}$ | Calculated $\mu_{eff}$ |
|---|---|---|---|
| $Tb_2Ti_2O_7$ | -14.75K | 9.5 $\mu_B$ | 9.29 $\mu_B$ |
| $Tb_{1.5}Eu_{0.5}Ti_2O_7$ | -19.8K | 8.41 $\mu_B$ | 8.66 $\mu_B$ |
| $Tb_{1.0}Eu_{1.0}Ti_2O_7$ | -24.12K | 7.12 $\mu_B$ | 7.313 $\mu_B$ |

Table (3) : Evaluated C.W. temperature, theoretical and calculated magnetic moment of $Tb_2Ti_2O_7$, $Tb_{1.5}Eu_{0.5}Ti_2O_7$, $Tb_{1.0}Eu_{1.0}Ti_2O_7$ samples from C.W. Fit ( 25-300 K) .

**(B)** Curie-Weiss Fit for samples $Tb_{0.2}Eu_{1.8}Ti_2O_7$, $Tb_{0.1}Eu_{1.9}Ti_2O_7$ in higher temperature range (150-300K) in which crystal field interactions are dominant. We have calculated the effective magnetic moment for these three samples whose values are listed in table (4).

| Sample | Theoretical $\mu_{eff}$ | Calculated $\mu_{eff}$ |
|---|---|---|
| $Tb_{0.2}Eu_{1.8}Ti_2O_7$ | 4.47$\mu_B$ | 5.25$\mu_B$ |
| $Tb_{0.1}Eu_{1.9}Ti_2O_7$ | 4.018$\mu_B$ | 4.704 $\mu_B$ |

Table (4): Evaluated theoretical and calculated magnetic moment of $Tb_{0.2}Eu_{1.8}Ti_2O_7$, $Tb_{0.1}Eu_{1.9}Ti_2O_7$ samples from C.W. ( 150-300 K) Fit.

**(C)** To find out the contribution of nearest neighbour exchange interactions energy ($J_{nn}$) and dipolar interaction energy ($D_{nn}$), We have fitted the data of Eu rich samples which show strange behaviour at low temperature by the high temperature series expansion of the susceptibility (by plotting $\chi T$ vs. 1/T) i.e.

$$\chi = C \left[ \left(\frac{1}{T}\right) + \left(\frac{\theta_{Cw}}{T^2}\right) \right] \qquad (4)$$

From the fit, we have extracted the values of Curie Weiss temperature ($\theta_{Cw}$), effective magnetic moment ($\mu_{eff}$), exchange interaction energy ($J_{nn}$) and dipolar interaction energy ($D_{nn}$). Here we have considered only classical interactions so we have calculated the classical exchange interaction energy ($J^{cl}$) using the relation $J^{cl} = S(S+1)J_{nn}$ and $J_{nn} = 3\theta_{Cw}/zS(S+1)$ [here z=6 is the co-ordination number]. The dipolar interaction energy is determined from $D_{nn} = \frac{\mu_0 \mu_{eff}^2}{4\pi r_{nn}^3}$ formula where $r_{nn}$ is the distance between a $R^{3+}$ ion at (000) and its nearest neighbour at (a/4, a/4,0), **a** being the lattice constant of the unit cell of the compound [47]. By using all these relations the calculated data are collected in table (5). From this table it is significant that on increasing the concentration of Eu ions, the classical exchange interaction energy is dominant to dipolar interaction energy which is the reason for enhancement of AFM nature and decrease in effective magnetic moment. The C.W. temperature is also more negative. All the related data of $Eu_2Ti_2O_7$ is consistent with earlier report [46]. The effective theoretical and calculated (by C.W. Fit) magnetic moment of all the compounds at room temperature are comparable to each other which is presented in the inset of Fig.4(b).

| Sample | $\theta_{CW}$ | $J^{cl}$ | $D_{nn}$ | Calculated $\mu_{eff}$ |
|---|---|---|---|---|
| $Tb_{0.2}Eu_{1.8}Ti_2O_7$ | -0.811K | -0.406K | 0.1179K | 2.98$\mu_B$ |
| $Tb_{0.1}Eu_{1.9}Ti_2O_7$ | -0.85K | -0.4275K | 0.04058K | 1.75$\mu_B$ |
| $Eu_2Ti_2O_7$ | -1.35K | -0.6742K | 0.00609K | 0.679$\mu_B$ |

Table (5) : Evaluated C.W. temperature, classical exchange energy, dipolar interaction energy and calculated magnetic moment of $Tb_{0.2}Eu_{1.8}Ti_2O_7$, $Tb_{0.1}Eu_{1.9}Ti_2O_7$, $Eu_2Ti_2O_7$ samples from High Temperature Series Expansion Fit [2-5K].

The variations of magnetization of all the series compounds with applied magnetic field at 2 K are presented in Fig. 4(b). M-H curves of TETO show unsaturated linear nature i.e. magnetic moment does not saturate up to 2 T applied magnetic field. This confirms the antiferromagnetic interaction and also supports the decrement of magnetic moment with increase of Eu content due to the enhancement of AFM exchange interactions.

*AC magnetic Study*

In order to investigate the the spin relaxation mechanism, AC susceptibility of all the samples were measured. In $Tb_2Ti_2O_7$ compound at H =0 Oe, real part of AC susceptibility $\chi'$ (T) (shown in fig. 6(a)) does not show any anomaly and shows canonical paramagnetic behaviour as earlier reported [60]. The imaginary part of the AC susceptibility at high temperature (inset (i) of Fig.6a) shows strange nature, where a relaxation peak at ~33K ($T_f$) is observed which is shifted towards high temperature with increasing frequency. This transition is thermally activated as it follows the Arrhenius law $f = f_0 e^{\frac{-E_a}{(k_B)(T_f)}}$ where $E_a$ is the activation energy for spin fluctuation and $f_0$ is a measure of the microscopic limiting frequency in the system and $k_B$ is the Boltzmann constant. The Arrhenius fit of TTO is shown in inset (ii) of Fig.6(a). As this transition is frequency dependent so to check the nature of the spin freezing, we have used the Mydosh formula p = $\frac{\Delta T_f}{T_f \Delta (log f)}$ , where $T_f$ is the freezing temperature at frequency f and p is the parameter which should be less than 0.01 for spin glass [16] transition. But for TTO this value is found to be 0.36 which is much larger than 0.01 value of spin glass. This value of p suggests that this transition is not spin glass type. Moreover, the fact that the magnetic field of 1T does not affect the $T_f$ also supports that the system is not a typical spin glass as for spin glass it is found that magnetic field suppresses the freezing temperature. The other pyrochlore systems like $Ho_2Sn_2O_7$ shows a spin freezing transition at low temperature [61] whereas $Ho_2Ti_2O_7$, $Dy_2Ti_2O_7$ show a spin freezing at low and high temperatures both [62-63]. Another parent compound ETO also shows a spin freezing transition below 35K [46]. However, the observation of this relatively higher temperature spin freezing for $Tb_2Ti_2O_7$ can be of particular interest as so far it was believed to remain dynamic down to 50 mK and no freezing was observed. Considering this spin freezing process in frustrated systems, we have analyzed the frequency distribution of spin relaxation mechanism of TTO pyrochlore in Fig. 6(b), by ploting the variation between $\chi''$ with frequency f. From this figure it can be seen that there is a consistent change in shape and pattern with increasing temperature as it reaches towards spin freezing transition temperature T~33K and shows a relatively sharp peak. For a single ion relaxation process, data can be interpreted by Casimir-du Pre relation which is $\chi''(f) = f\tau[(\chi_T - \chi_S)/(1+f^2\tau^2)]$, where $\chi_T$ is isothermal susceptibility in the limit of low frequency and $\chi_S$ is the adiabatic susceptibility in the limit of high frequency [64]. From inset of Fig. 6(b), it can be seen that for temperature 33.1K and near this temperature, although the normalized $\chi''$ fits with the relation, quality of that fitting i.e. Cole -Cole (Argand plot ) of "$\chi'$ **Vs** $\chi''$", does not show expected semicircle (not shown) [65]. This behaviour of Argand plot is possibly due to the frequency independent value of $\chi'$ at spin freezing temperature. From all these analyses and facts, it is confirmed that this glassy behaviour in $Tb_2Ti_2O_7$ is not spin glass type, but it can be termed as a kind of weak spin freezing. The reason behind $\chi'$ curve not showing anomaly near freezing temperature or signature of spin freezing could be a faster spin relaxtion in $\chi'(T)$ curve of TTO.

To find out the nature of this weak spin freezing, i.e. whether this spin freezing is associated with spin-spin correlation or some other interaction, we analyzed the effect of non-magnetic dilution (by substituting Y on Tb site) in TTO system. For this, we have measured the ac magnetic property of $Y_{1.8}Tb_{0.2}Ti_2O_7$ (YTTO) sample. In this sample, we have substituted 90% of Tb ions by non magnetic Y ions so that $Y^{3+}$ can change the local environment of $Tb^{3+}$ ions. Fig. 7(a) shows the ac magnetization of both YTTO and pure TTO. It can be clearly seen that in $\chi''(T)$ graph [inset of Fig.7(a)], non magnetic dilution enhances the spin freezing instead of suppressing it. For spin spin correlation, this spin freezing should be diminished. This result confirms the single ion nature of spin freezing in TTO compound. This kind of single ion spin freezing is also reported in $Eu_2Ti_2O_7$ [46] and in $Dy_2Ti_2O_7$ [66].

It is quite interesting to see the involvement of $Eu^{3+}$ ions on this spin relaxation mechanism or spin freezing of TTO compound as TETO series samples also show a single ion spin freezing at about (~32K for 500 Hz frequency). Here we have analyzed the AC magnetic susceptibility of all the TETO compounds. In Fig. (8a) and (8b) we plotted the temperature dependence $\chi'$ and $\chi''$ curve for all the systems at zero magnetic field (H = 0 Oe) for particular 500 Hz frequency. In Fig. (8 a), $\chi'$ of all the compounds shows a canonical paramagnetic behaviour except ETO. $\chi'(T)$ of TETO decreases continuously with increasing $Eu^{3+}$ content. ETO compound shows appearance of a clear dip around 35 K in $\chi'(T)$ [inset of Fig. 8(a)] and the imaginary part of ac susceptibility corresponding to the dip in $\chi'(T)$ shows a clear spin freezing transition below 35K ($T_f$) (inset Fig.8 (b)). This result of ETO compound is consistent with earlier reported data [46]. In Fig. 8(b), the imaginary part of susceptibility of all the samples can be seen to show a single ion spin freezing below 35 K. Actually TTO shows a weak spin freezing at ~ 33K (500 Hz) and ETO shows a single ion spin freezing at ~32K (500 Hz) so it is quite difficult to differentiate between the spin freezing temperature of individual Tb and Eu ions for TETO samples. We have also shown the Arrhenius fit of TETO compounds in Fig 7(b) as all the samples shows thermally activated mechanism. From the Arrhenius fit, we have extracted the value of $E_a$ (thermal energy barrier) and $f_0$ (characteristic frequency) for all series samples. The value of $E_a$ are 198.65 K, 182.946 K, 184.94 K, 186.30 K for $\{Tb_{1-x}Eu_x\}_2Ti_2O_7$ with (x= 0.25,0.50,0.90,0.95) respectively and value of $f_0$ for all are in range of ~$10^5$ Hz. This relatively high value of characteristic relaxation time suggests towards slow spin relaxation occurring near the freezing temperature. The change in $E_a$ values arise in the samples as result of the small alteration in the crystal Field levels as the lattice constant is increased on increase of Eu concentration.

To observe the effect of magnetic field on the AC susceptibility, a field of 1T was applied for all the samples. Fig.(9) shows the ac susceptibility data of all the samples at a DC magnetic field of 1T. In Fig 9(a) for TTO sample, a clear dip appears around at low temperature ~ 4K ($T^*$) in the real part of susceptibility corresponding to Kramers - Kronig relation. A sharp rise ~ 4K also appears at imaginary part of susceptibility measurement. A peak at ~33K ($T_f$) associated to weak spin freezing is also present in the $\chi^{//}(T)$ part. This field induced transition is due to the single moment saturation process as earlier reported in TTO [61]. The same scenario is found for all other samples as field induced transition ($T^*$) appears at lower temperature side and spin freezing transition ($T_f$) found at higher temperature side which is plotted in Fig. 9(b-e) respectively. The ($T^*$) peak systematically shifts towards higher temperature with increasing the concentration of Eu ions. The peak positions of ($T^*$) are listed in a table (6)

| x | 0 | 0.25 | 0.50 | 0.90 | 0.95 | 1.0 |
|---|---|------|------|------|------|-----|
| Temperature | ~ 4.7K | ~ 6.15K | ~ 6.8K | ~ 7.59K | ~ 8.13K | - |

<u>Table (6):</u>-Positions of field induced ($T^*$) peak for $\{Tb_{1-x}Eu_x\}_2Ti_2O_7$ with (x= 0,0.25,0.50,0.90,0.95,1.0)

The shifting of ($T^*$) peak with $Eu^{3+}$ doping is possibly due to a systematic change in its structure and analogous variation in the crystal field levels as the crystal field depends on the different structural parameters like, lattice constants and positional parameters of $R^{+3}$ ions [44]. For ETO sample, we did not found any field induced transition which is consistent with earlier report data [Fig.9(f)] [46]. This ($T^*$) peak of Eu doped TTO samples is independent of the frequency so it is not related to thermal relaxation process of spins. Not being thermally activated they also did not follow Arrhenius Law. The nature of this ($T^*$) peak for TETO samples is similar to the ($T^*$) peak of pure TTO samples so this new field induced transition may be associated with single moment saturation [60]. The surprising point in this measurement is that the position of the freezing ($T_f$) peaks remained almost unaltered i.e. positioned below ~34 K for all TETO samples. There is no change in its position even in presence of magnetic field of 1T except a little suppression in the spin freezing which again confirms its single ion spin freezing nature. For $Tb_{0.2}Eu_{1.8}Ti_2O_7$ and $Tb_{0.1}Eu_{1.9}Ti_2O_7$ samples, the spin freezing was observed even in the $\chi^/(T)$ curves which possibly indicate towards relatively slower spin relaxation associated to the $Eu^{3+}$ spins. The behaviour of $(T^*)$ peak in $\chi''(T)$ curve for last two samples of TETO series i.e. for $Tb_{0.2}Eu_{1.8}Ti_2O_7$ and $Tb_{0.1}Eu_{1.9}Ti_2O_7$ is little different from first three samples i.e. $Tb_2Ti_2O_7$, $Tb_{1.5}Eu_{0.5}Ti_2O_7$, $Tb_{1.0}Eu_{1.0}Ti_2O_7$. As for first three samples $(T^*)$ peak show a monotonous growth but the shape of peak is not clear but for last two samples, this low temperature field induced transition peak is quite prominent. The reason for that strange behaviour may be the more interaction between Tb-Eu ions as compared to interaction between Tb-Tb ions. To fully understand the origin of this new spin freezing in TTO Pyrochlore and to know the unrevealed concepts of spin mechanism, more measurement like neutron diffraction and specific heat may help us.

## *Conclusion*

We have systematically studied the structural, Raman and magnetic properties of $(Tb_{1-x}Eu_x)_2Ti_2O_7$ (with x= 0,0.25,0.50,0.90,0.95,1.0) pyrochlore series. The formation of all the compositions and the systematic substitution of $Eu^{3+}$ ions in place of $Tb^{3+}$ ions are analyzed from its structural and Raman Studies. DC magnetization study shows the enhancement of antiferromagnetic interaction and decrease in total magnetic moment with Eu doping. AC magnetic susceptibility study of particular $Tb_2Ti_2O_7$ shows a new single ion weak spin freezing ~ 33K at zero magnetic field. For rest of the samples similar spin freezing is found. On applying a field of 1T, all the compounds show a new field induced transition $(T^*)$ which shifts towards higher temperature side on increasing the concentration of Eu ions. This field induced transition peak corresponds to single moment saturation and the $(T_f)$ peak of spin freezing also appeared in imaginary part of AC susceptibility measurement. However these results suggest that it deserves further study to find out the origin of this weak spin freezing of TTO sample.

## *Acknowledgements*

The authors are thankful to the CIF, Indian Institute of Technology (BHU) for providing the facility of magnetic measurements. PS and VKG are also expressing their gratitude to UGC India for providing fellowship. The financial support by DST-FIST to the department is also greatfully acknowledged.

## *Table Caption :-*

Table (1): The ionic radius ratio of TETO series compounds.

Table (2) : Different parameters obtained from rietveld refinement of all TETO series compounds with Space Group F d -3 m .

Table (3) : Evaluated C.W. temperature, theoretical and calculated magnetic moment of $Tb_2Ti_2O_7$, $Tb_{1.5}Eu_{0.5}Ti_2O_7$, $Tb_{1.0}Eu_{1.0}Ti_2O_7$ samples from C.W. Fit ( 25-300 K) .

Table (4): Evaluated theoretical and calculated magnetic moment of $Tb_{0.2}Eu_{1.8}Ti_2O_7$, $Tb_{0.1}Eu_{1.9}Ti_2O_7$ samples from C.W. ( 150-300 K) Fit.

Table (5) : Evaluated C.W. temperature, classical exchange energy, dipolar interaction energy and calculated magnetic moment of $Tb_{0.2}Eu_{1.8}Ti_2O_7$, $Tb_{0.1}Eu_{1.9}Ti_2O_7$, $Eu_2Ti_2O_7$ samples from High Temperature Series Expansion Fit [2-5K].

Table (6):- Positions of field induced ($T^*$) peak for $\{Tb_{1-x}Eu_x\}_2Ti_2O_7$ with (x= 0,0.25,0.50,0.90,0.95,1.0)

## *Figure Caption :-*

Fig.(1): Powder x-ray diffraction pattern for the $(Tb_{1-x}Eu_x)_2Ti_2O_7$ samples. The inset shows the diffraction peak (222) of $(Tb_{1-x}Eu_x)_2Ti_2O_7$ samples.

Fig.(2)- Rietveld refinement for the $(Tb_{1-x}Eu_x)_2Ti_2O_7$ samples.

Fig.3(a) Raman Spectra of the $(Tb_{1-x}Eu_x)_2Ti_2O_7$ samples at 300K. (b) Variation of all active phonon modes as a function of x with the straight horizontal lines for reference.

Fig.4(a) The dc susceptibility of the $(Tb_{1-x}Eu_x)_2Ti_2O_7$ samples at ZFC. The inset shows the inverse dc susceptibility of $(Tb_{1-x}Eu_x)_2Ti_2O_7$ samples.

Fig.4 (b) The magnetization Vs magnetic field at 2K for all TETO series samples. The inset shows the effective magnetic moments derived from C.W. Law at high temperature for TETO series samples.

Fig.5(a) The Curie – Weiss Fit for $Tb_2Ti_2O_7$, $Tb_{1.5}Eu_{0.5}Ti_2O_7$, $Tb_{1.0}Eu_{1.0}Ti_2O_7$ samples .

Fig.5(b) The High Temperature Series Expansion Fit for $Tb_{0.2}Eu_{1.8}Ti_2O_7$, $Tb_{0.1}Eu_{1.9}Ti_2O_7$, $Eu_2Ti_2O_7$ samples.

Fig.6(a) The real part ($\chi'$) of ac susceptibility of $Tb_2Ti_2O_7$ as a function of temperature at zero applied DC field. The inset (i) shows the imaginary part ($\chi''$) of ac susceptibility of $Tb_2Ti_2O_7$. The inset (ii) shows Arrhenius Fit of $Tb_2Ti_2O_7$.

Fig.6(b) Variation of ($\chi''$) imaginary part of ac susceptibility as a function of frequency at different temperatures for $Tb_2Ti_2O_7$. The inset shows the normalized $\chi''$ as a function of $f/f_{peak}$ and fitted theoretically (red) by Casimir du pre relations at and near the transition temperature for $Tb_2Ti_2O_7$.

Fig.7(a) The real part ($\chi'$) of ac susceptibility of TTO and YTTO compounds at H = 0 Oe. The inset shows the imaginary part ($\chi''$) of ac susceptibility of TTO and YTTO compounds at H = 0 Oe.

Fig.7(b) The Arrhenius Fit of ($T_f$) peak for compounds.

Fig. 8 (a) The variation of real part ($\chi'$) of ac susceptibility of all TETO series compounds as a function of temperature for frequency of 500 Hz at zero applied DC field. The inset shows the real part ($\chi'$) of ac susceptibility of ETO sample at H = 0 Oe.

Fig.8 (b) The Variation of imaginary part ($\chi''$) of ac susceptibility of all TETO series compounds as a function of temperature for frequency of 500 Hz at zero applied DC field. The inset shows imaginary part ($\chi''$) of ac susceptibility of ETO sample at H = 0 Oe.

Fig. 9 The variation of real part ($\chi'$) and imaginary part ($\chi''$) of ac susceptibility of all TETO series compounds as a function of temperature at applied field of 10 k Oe.(a) $Tb_2Ti_2O_7$ (b) $Tb_{1.5}Eu_{0.5}Ti_2O_7$ (c) $Tb_{1.0}Eu_{1.0}Ti_2O_7$ (d) $Tb_{0.2}Eu_{1.8}Ti_2O_7$ (e) $Tb_{0.1}Eu_{1.9}Ti_2O_7$ (f) $Eu_2Ti_2O_7$. Marked by arrow are : Single momemt saturation peak ($T^*$) and the Single ion spin freezing peak ($T_f$).

## *Figures*

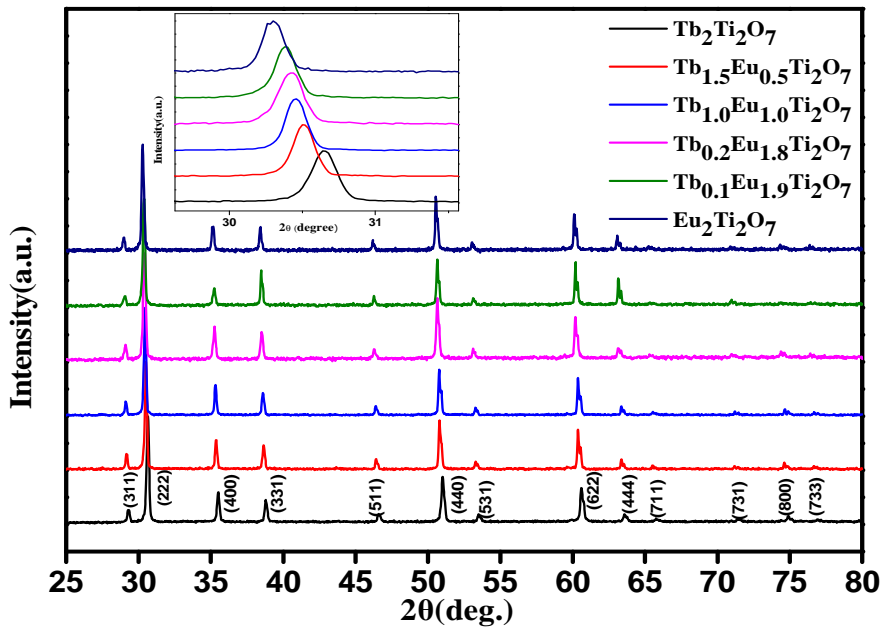

Fig.(1)

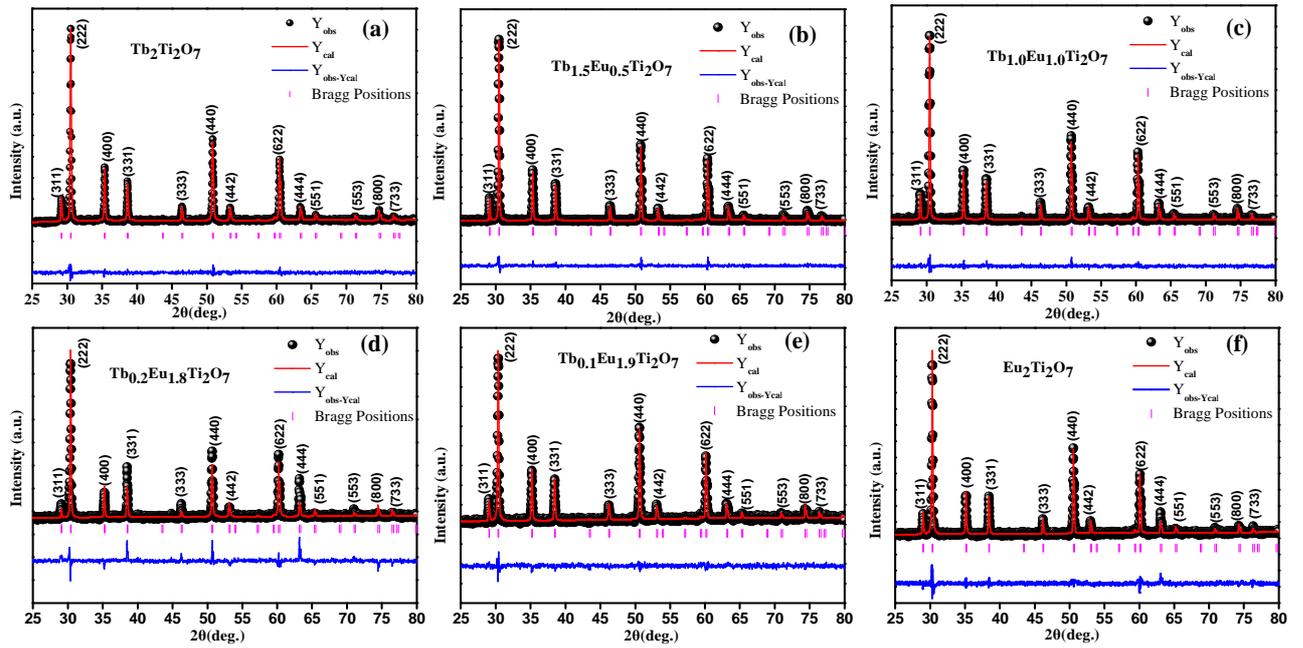

Fig.(2)

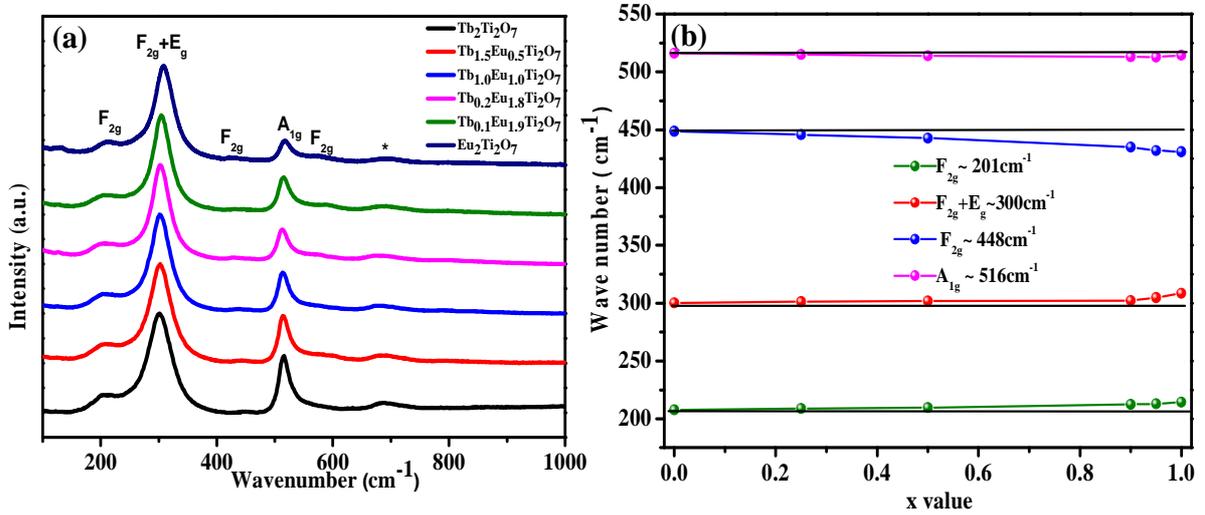

Fig.(3)

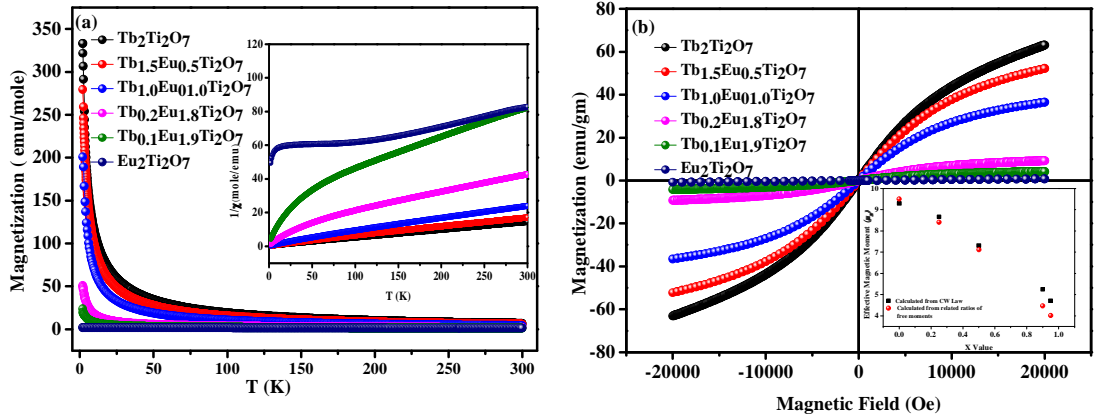

Fig.(4)

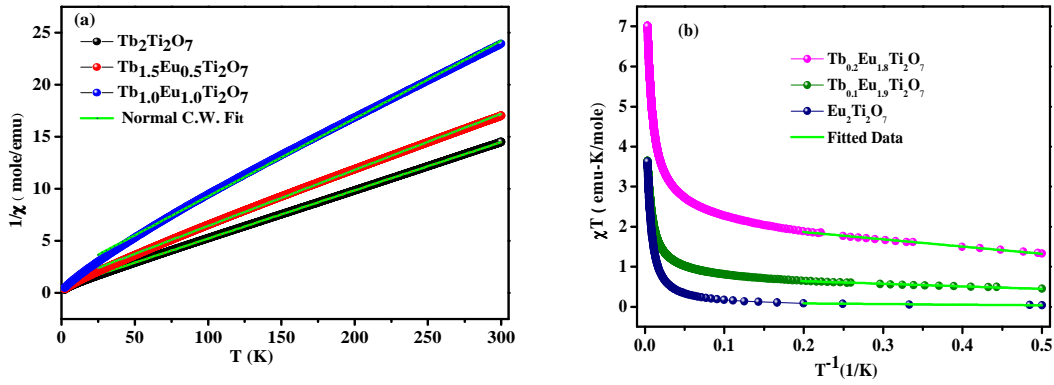

Fig.(5)

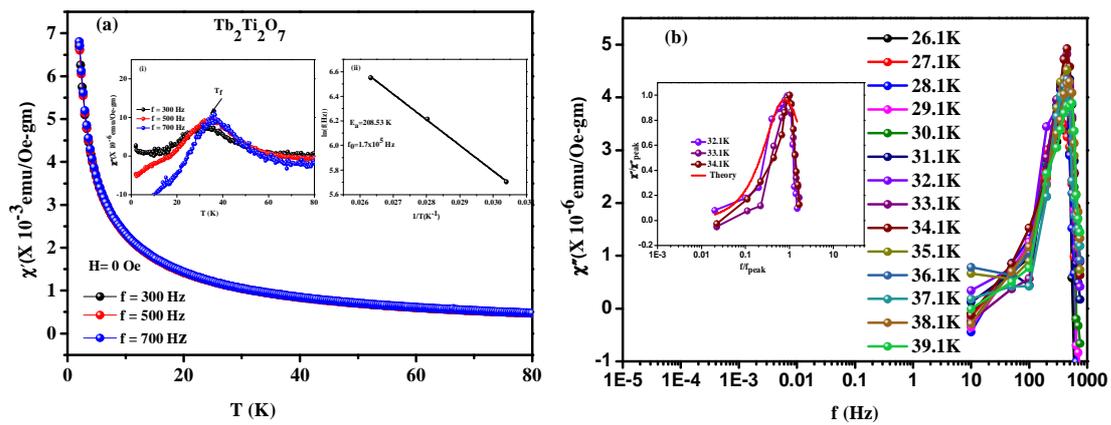

Fig.(6)

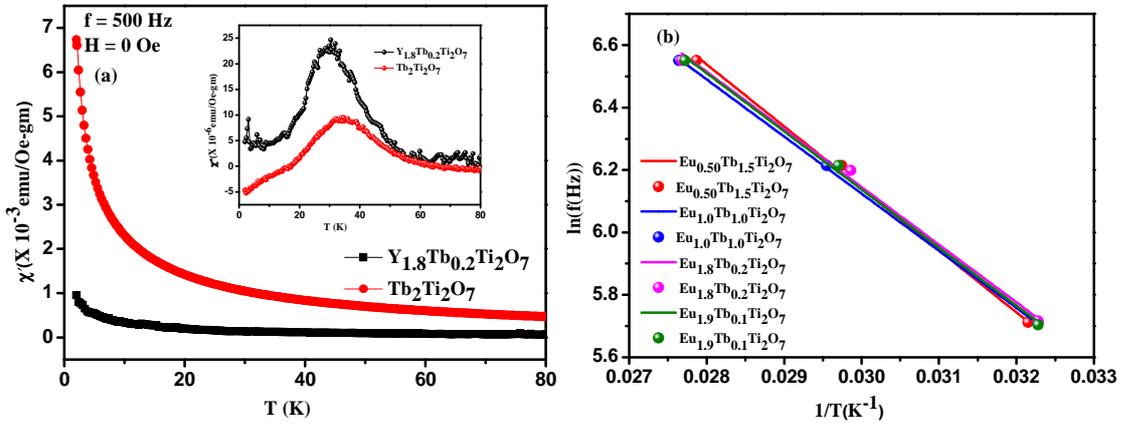

Fig.(7)

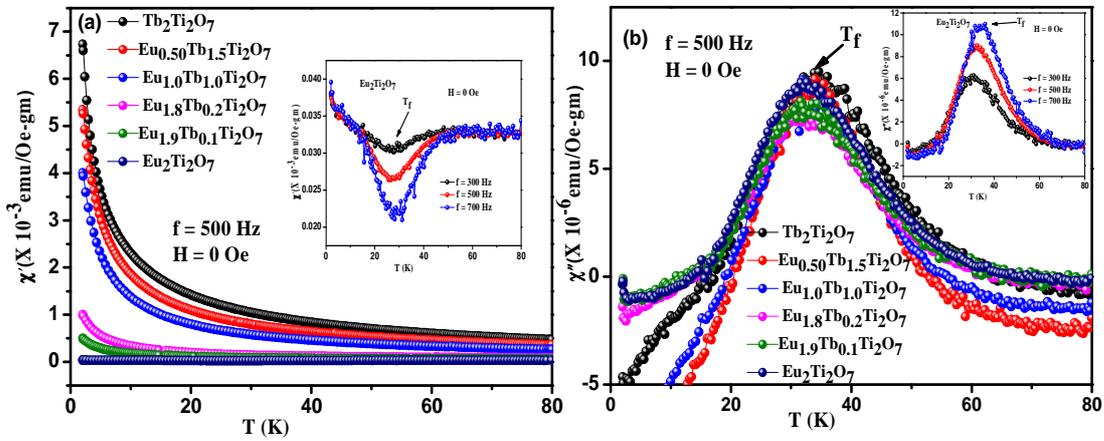

Fig. (8)

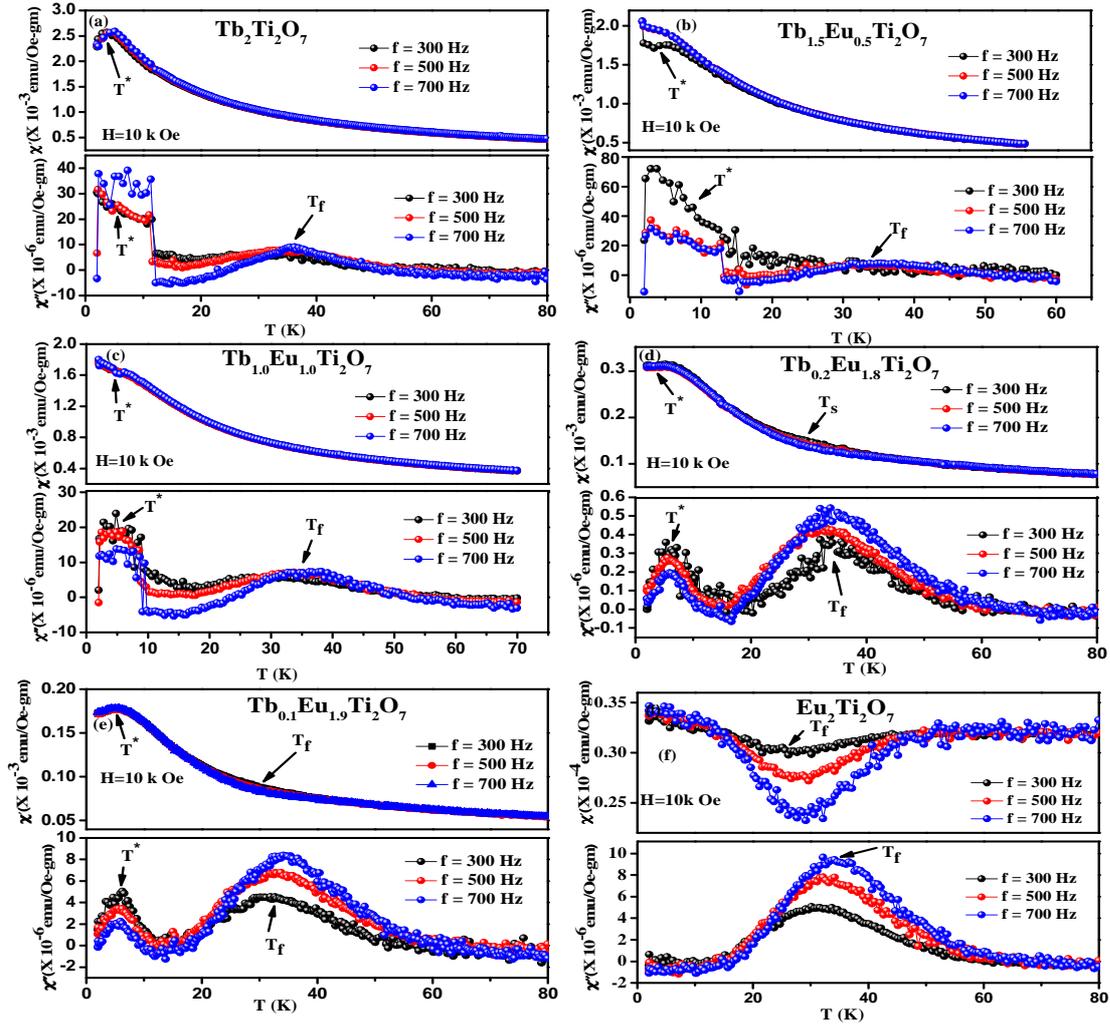

Fig. (9)